\documentclass[sigconf]{acmart}

\AtBeginDocument{%
  }

\copyrightyear{2025}
\acmYear{2025}
\setcopyright{cc}
\setcctype{by}
\acmConference[SIGIR '25]{Proceedings of the 48th International ACM SIGIR Conference on Research and Development in Information Retrieval}{July 13--18, 2025}{Padua, Italy}
\acmBooktitle{Proceedings of the 48th International ACM SIGIR Conference on Research and Development in Information Retrieval (SIGIR '25), July 13--18, 2025, Padua, Italy}\acmDOI{10.1145/3726302.3730229}
\acmISBN{979-8-4007-1592-1/2025/07}



\usepackage[T1]{fontenc}
\usepackage{graphicx}

\usepackage[show]{chato-notes}
\usepackage{booktabs}
\usepackage{pifont}
\usepackage{multirow}
\usepackage{lipsum}
\usepackage{subcaption}
\usepackage{amsmath}
\usepackage{textcomp}
\usepackage{xcolor}
\usepackage{tcolorbox}
\usepackage{fontawesome5}

\newcommand{\precone}{\ding{172}P}
\newcommand{\recone}{\ding{172}R}
\newcommand{\prectwo}{\ding{173}P}
\newcommand{\rectwo}{\ding{173}R}

\newcommand{\jm}[1]{\textcolor{black}{#1}}
\newcommand{\gm}[1]{\textcolor{black}{#1}}

\newcommand{\crc}[1]{\textcolor{black}{#1}}

\newcommand{\observation}[1]{%
\begin{tcolorbox}[colback=blue!10!white, colframe=blue!10!white, sharp corners]
  #1
\end{tcolorbox}
}

\usepackage{marginnote}
\newcommand{\pageenlarge}[1]{\marginnote{}\enlargethispage{#1\baselineskip}}

\begin{document}

\title{Measuring Hypothesis Testing Errors in the Evaluation of Retrieval Systems}

\author{Jack M\textsuperscript{c}Kechnie}
\affiliation{%
  \institution{University of Glasgow}
  \city{Glasgow}
  \country{United Kingdom}}
\email{j.mckechnie.1@research.gla.ac.uk}

\author{Graham McDonald}
\affiliation{%
  \institution{University of Glasgow}
  \city{Glasgow}
  \country{United Kingdom}}
\email{graham.mcdonald@glasgow.ac.uk}

\author{Craig Macdonald}
\affiliation{%
  \institution{University of Glasgow}
  \city{Glasgow}
  \country{United Kingdom}}
\email{craig.macdonald@glasgow.ac.uk}

\renewcommand{\shortauthors}{Jack McKechnie, Graham McDonald, \& Craig Macdonald}

\begin{abstract}
\looseness -1
\pageenlarge{2}
The evaluation of Information Retrieval (IR) systems typically uses query-document pairs with corresponding human-labelled relevance assessments (qrels). These qrels are used to determine if one system is better than another based on average retrieval performance. \jm{Acquiring large volumes of human relevance assessments is expensive. Therefore, more efficient relevance assessment approaches have been proposed, necessitating comparisons between qrels to ascertain their efficacy.} Discriminative power\jm{,} i.e. the ability to correctly identify significant differences between systems, is important for drawing accurate conclusions on the robustness of qrels. Previous work has measured the proportion of pairs of systems that are identified as significantly different and has quantified Type I statistical errors. Type I errors lead to incorrect conclusions due to false positive significance tests. We argue that also identifying Type II errors (false negatives) is important as they lead science in the wrong direction. We quantify Type II errors and propose that balanced classification metrics\jm{,} such as balanced accuracy\jm{,} can be used to portray the discriminative power of qrels. We perform experiments using \jm{qrels generated using alternative relevance assessment methods} to investigate measuring hypothesis testing errors in IR evaluation. We find that additional insights into the discriminative power of qrels can be gained by quantifying Type II errors, and that balanced classification metrics can be used to give a\jm{n overall} summary of discriminative power \jm{in one, easily comparable, number}.
\end{abstract}


\maketitle

\section{Introduction}
The offline evaluation of Information Retrieval (IR) systems commonly follows the Cranfield paradigm~\cite{cranfield}, using test collections and evaluation measures to determine the quality of the results of an IR system. Such test collections comprise a corpus of documents, a set of queries, and query-relevance assessments (qrels). Qrels are typically constructed by performing top-$k$ pooling, where top-ranked documents from several retrieval systems are manually judged for relevance.\pageenlarge{2} Qrels are used to quantify system performance using an evaluation measure. Consequently, systems can be compared and statistical testing can be \jm{performed} on their respective results to inform scientific conclusions and deployment decisions.

\looseness -1
Test collections are crucial in the development of new IR systems. However, collecting \jm{expert} human relevance assessments is expensive. This high cost can prohibit new test collections from being built~\cite{one_shot_labelling,too_expensive}, leading to investigations into the use of alternative assessors such as crowdworkers~\cite{crowdworkers} and Large Language Models (LLMs)~\cite{llms_fill,perspectives,llm4eval,llms_patch,ecir_context}. \jm{Furthermore, to reduce assessment costs and the required annotator time, document prioritisation methods~\cite{mtf_pooling,bandit_pooling,ntcir_pooling,top_k_pooling} have been developed.} To determine the quality of the qrels produced by such approaches (\emph{candidate qrels}), we must compare the resulting qrels against those that we believe to be of high quality (\emph{ground truth qrels}), typically human judgements over top-$k$ pools~\cite{auto_eval_soboroff,thomas_predict}. \crc{In this work, we take human judgements as our ground truth. Human relevance judgements are the gold standard in IR evaluation~\cite{relevance_def_1,relevance_def_2,relevance_def_3} and consequently we must compare against such labels}. Such comparisons require the use of appropriate metrics to ensure that valid conclusions are drawn.

\begin{figure}[tb]
    \centering
    \includegraphics[width=0.8\linewidth]{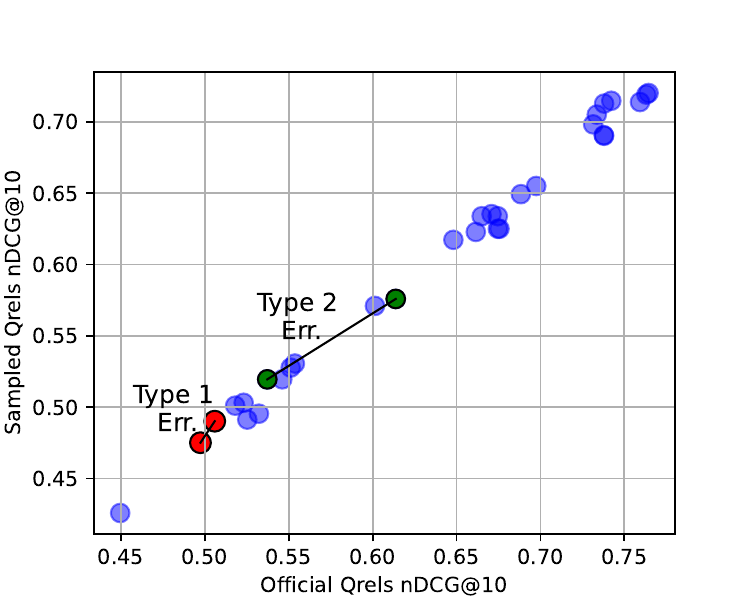}  \vspace{-4mm}
    \caption{\looseness -1 Illustration of hypothesis testing errors. $\tau$ 0.9\crc{2} $\Delta$ Sensitivity 0. Examples of \crc{Type} I and II errors that occur when evaluating with this set of candidate qrels are highlighted.}
    \label{fig:delta_zero}
    \vspace{-6mm}
\end{figure}

\looseness -1
Previous work comparing qrels~\cite{llms_fill,auto_eval_soboroff,thomas_predict,llms_patch} has used agreement measures such as Cohen's $\kappa$~\cite{cohen_kappa} to quantify the agreement of relevance labels and ranking correlation measures such as Kendall's $\tau$~\cite{kendall_tau} to compare the ranking of systems by an evaluation measure. However, \emph{discriminative power}\jm{,} i.e. how well qrels can identify significant differences between systems, is important to measure since we want to know which systems are significantly better than other systems. To this end, Faggioli et al.~\cite{perspectives} proposed the sensitivity metric. Given retrieval results from several systems, the sensitivity metric measures the proportion of all pairs of systems that are identified as significantly different. Sensitivity is calculated using ground truth qrels and candidate qrels usually qrels gathered in a novel manner. \pageenlarge{2}Sensitivity \jm{is} reported as the difference ($\Delta$) in the number of significant differences between systems, as measured by each of the qrels. However, we argue that measuring discriminative power \jm{using sensitivity} can obscure hypothesis-testing errors as \jm{the sensitivity measure only considers if pairs of systems are significantly different according to the candidate qrels, not to the ground truth qrels.} \jm{The candidate qrels may not be able to correctly identify all of the significant differences between systems that the ground truth qrels are able to identify. Thus, measuring the discriminative power of qrels is important as it allows us to quantify the extent to which the candidate qrels are able to correctly identify significant differences.} Figure~\ref{fig:delta_zero} shows nDCG@10 according to ground truth qrels on the x-axis and according to candidate qrels on the y-axis, with points representing systems. This set of candidate qrels results in $\Delta$ sensitivity of 0, the ground truth and candidate qrels identify the same number of significant differences. One may therefore conclude that the candidate qrels are as discriminative as the ground truth qrels. However, the candidate qrels identify pairs of systems as being significantly different when they, in fact, are not; and identify pairs of systems as being not significantly different when they are, according to the ground truth qrels. Such errors are known as Type I and Type II errors respectively. 
\crc{Using these candidate qrels results in 10 Type I and 10 Type II errors. Figure~\ref{fig:delta_zero} presents an illustrative example of each of these types of errors.}


Consequently, measuring discriminative power using only $\Delta$ sensitivity can lead to incorrect conclusions since Type I and II errors can cancel each other out. On the other hand, Otero et al.~\cite{how_discrim} proposed to measure discriminative power using classification metrics. The outcome of a significance test is treated as a binary label of (non) significance. Testing for significance using ground truth and candidate qrels, Otero et al.~\cite{how_discrim} measured discriminative power using precision and recall metrics. Here, precision measures the correctness of the \jm{significant differences identified by the candidate qrels compared to the significant differences identified by the ground truth qrels and} recall measures \jm{the} coverage. In this sense, high precision indicates low Type I errors as few significant differences are incorrectly identified. Whilst the methods proposed by Otero et al\jm{.}~\cite{how_discrim} consider Type I errors, they do not consider correctly identifying \emph{non-significant} differences and consequently disregard Type II errors. We argue that it is also important to quantify Type II errors. Committing Type II errors can lead science astray; effective systems that ultimately should be deployed and further investigated are discarded as they are falsely concluded to be no better than others~\cite{waste_time_money,type2_errors_bad}.

\looseness -1
\jm{We seek to measure the Type II errors that occur when using different qrels to evaluate retrieval systems. Indeed, other areas of \crc{I}nformation \crc{R}etrieval evaluation are also affected by hypothesis testing errors. For example, in recommender systems research, datasets are split into train and test sets. How the data is split affects the evaluation outcomes~\cite{fresh_look_rs,item_sampling_rs,data_splitting_rs,comparing_frameworks_rs}, which is well documented in recommended systems evaluation literature~\cite{movielens_different_outcomes}. We argue that measuring the hypothesis testing errors that occur when using different data splits would also additional insights into the performance of recommender systems.}

\looseness -1
\jm{In this paper we argue that when evaluating the robustness of candidate qrels it is important to measure the precision and recall of the \emph{non-significant} differences between IR systems that the candidate qrels are able to correctly identify. The precision of non-significant differences being identified enables us to measure how many of the differences between systems identified as non-significant truly are non-significant according to the ground-truth qrels.} Recall of non-significant differences being identified allows us to measure how many of the truly non-significant differences are identified. \jm{Committing Type II errors can lead science astray as we falsely conclude that there is no difference between two systems. When developing novel methods for creating qrels, the risk of making such incorrect conclusions can be minimised by comparing Type II error rates with a set of ground truth qrels.} \pageenlarge{2}The combination of both precision and recall of non-significant differences allows us to draw further conclusions about the discriminative power of qrels \jm{in a manner that makes the rate of Type II errors clear}. Additionally, we propose to use \jm{balanced} classification metrics that consider both Type I and Type II errors to provide a \jm{distilled overview of the overall discriminative power of a set of candidate qrels. While these measures are intuitive, we are not aware of their previous use in the literature for measuring qrel quality.}

\vspace{-4mm}
\section{Related Work}

\looseness -1
\noindent\textbf{Statistical Testing in IR Experiments.} Significance testing in IR and the hypothesis testing errors that can occur have seen much study~\cite{testing_anova,testing_effect,test_subcorpora,testing_shards,testing_two,testing_optimality}. Zobel~\cite{zobel_topic_splitting} proposed a topic-splitting approach to measure the number of Type I errors that occurred when comparing runs from TREC-5~\cite{trec5} that was further investigated by Ferro and Sanderson~\cite{sanderson_test_a_test}. Such work on topic splitting finds that Type I errors often occur when evaluating IR systems. Fuhr~\cite{fuhr_mistakes} and Sakai~\cite{sakai_guidelines} advocated for the use of multiple testing correction to mitigate the chance of Type I errors, details of which are investigated by Boystov et al.~\cite{mutiple_correction}. However, rather than investigating how significance testing can be performed in IR experiments, we investigate how the errors in such significance testing can be quantified.

\noindent\textbf{Development of Alternative Qrels.} The cost of human relevance assessment~\cite{assessment_expensive} has prompted investigations into alternative methods for developing relevance assessments. Crowdworkers~\cite{crowdworkers} have been used as they are often paid at a low rate~\cite{crowdworkers_cheap}. Alternatively, efforts have gone into developing qrels using statistics from the pools created by a diverse set of retrieval systems~\cite{auto_eval_aslam,auto_eval_hauff,auto_eval_soboroff,auto_eval_wu}, however, such approaches assign relevance labels based on the popularity of retrieval rather than semantic relevance. We use this as a `strawman' baseline. Most recently, LLMs have been used to create relevance assessments~\cite{llms_fill,perspectives,llm4eval,llms_patch}.

\looseness -1
\noindent\textbf{Comparison of Qrels} Qrels have been compared in several ways. The agreement of relevance labels has been measured with Cohen's $\kappa$~\cite{llms_fill,cohen_kappa,llm4eval} and the correlation of the ranking of a set of systems~\cite{perspectives,auto_eval_soboroff,umbrella} has been measured using Kendall's $\tau$~\cite{kendall_tau} and Spearman's $\rho$~\cite{spearman_rho}. However, identifying which systems are statistically significantly different to others is important. To this end, Otero et al.~\cite{how_discrim} proposed to use precision and recall of the outcomes of significance tests between pairs of systems to determine the discriminative power of qrels. Finally, Faggioli et al.~\cite{perspectives} proposed another metric of discriminative power, namely sensitivity. Sensitivity~\cite{perspectives} measures the number of significant differences between pairs of systems. However, sensitivity does not account for hypothesis testing errors. Moreover, only Type I errors are considered by Otero et al.~\cite{how_discrim}. We therefore investigate quantifying Type II errors and how to report the overall discriminative power of qrels in a single metric.

\section{Measuring Hypothesis Testing Errors}
\jm{This section outlines the metrics that we propose to use for measuring Type II errors, the precision and recall metrics for measuring Type I errors proposed by Otero et al.~\cite{how_discrim}, and how the candidate qrels that we use for our comparisons are obtained.}

\looseness -1
\jm{To quantify hypothesis testing errors we cast the task of significance testing as a binary classification task. We define four sets; $S_{gt}$, $S_{\crc{cand.}}$, $NS_{gt}$, and $NS_{cand.}$. \jm{Significant differences between pairs of systems identified by the ground truth qrels are contained in $S_{gt}$ and by the candidate qrels in $S_{cand.}$. Non-significant differences between pairs of systems identified by the ground truth qrels are contained in $NS_{gt}$ and by the candidate qrels in $NS_{cand.}$.}} \jm{We then define the precision and recall metrics for measuring Type II errors as:}

\begin{center}
    \prectwo{} = $\frac{|NS_{gt} \cap NS_{cand.}|}{|NS_{cand.}|}$
    \rectwo{} = $\frac{|NS_{gt} \cap NS_{cand.}|}{|NS_{gt}|}$\\
\end{center}
\pageenlarge{2}
\noindent\jm{High \prectwo{} indicates that few Type II errors are made by the candidate qrels and high \rectwo{} indicates coverage of non-significant differences.}

\looseness -1
\noindent\textbf{Measuring Type I Errors} Otero et al.~\cite{how_discrim} proposed to use the precision and recall of significant differences being identified (\precone{} and \recone{}) to assess the discriminative power of qrels. \jm{Using the notation defined above, we can define \precone{} and \recone{} as:}

\begin{center}
    \precone{} = $\frac{|S_{gt} \cap S_{cand.}|}{|S_{cand.}|}$ 
    \recone{} = $\frac{|S_{gt} \cap S_{cand.}|}{|S_{gt}|}$\\
\end{center}

However, we argue that \emph{non-significance} is an important aspect of IR evaluation and, as such, the power of qrels to identify non-significant differences should also be measured. Therefore, we additionally report the precision (\prectwo{}) and recall (\rectwo{}) of non-significant differences being correctly identified. Such Type II errors (false negatives i.e. significant differences not being identified) lead science in the wrong direction. We also report Balanced Accuracy (BAC) and Matthews Correlation Coefficient (MCC) as these are balanced metrics that consider both Type I and Type II errors. We investigate if balanced classification metrics can provide a more rounded description of discriminative power.

\noindent\textbf{\crc{C}andidate qrels For Comparisons.} We now describe the three sets of qrels that we use to make our comparisons; Zero-Shot LLM, Percentage Sampling, Popularity-Biased Labeller.

\noindent\textit{\textbf{Zero-Shot \jm{LLM}}} This approach for creating qrels leverages an LLM that is prompted to generate relevance assessments in a zero-shot manner, i.e. no demonstration examples are \crc{used}. We instantiate this approach with the L\crc{l}ama 3 model~\cite{llama3} and a prompt based on Thomas et al.~\cite{thomas_predict}. The zero-shot approach is intended to represent the body of literature that uses LLMs for relevance assessment. We expect that this approach will rank systems well and that the semantic understanding of the LLM will result in a discriminative set of qrels.

\noindent\textit{\textbf{Percentage Sampling}} Similar to Otero et al.~\cite{how_discrim}\jm{,} we investigate using subsamples of the original qrels. We randomly sample the relevant documents in the original qrels. We expect that only a small percentage of all qrels will be necessary to accurately identify significantly different pairs of systems, as per~\cite{how_discrim}.

\noindent\textit{\textbf{Popularity-Biased Labeller}} To represent the family of work that uses the statistics from the pools used for assessment to generate relevance labels, we use the popularity-biased labeller approach~\cite{auto_eval_soboroff}. We label the $p$ documents that are most commonly retrieved by the systems used to create the pools as relevant and all other documents as non-relevant, where $p$ is the percentage of documents that are labelled relevant by the original qrels. We do not expect this approach to perform well in terms of system ranking and discriminative power as relevance labels are decided by the popularity of being retrieved\crc{,} not by semantic relevance.

\begin{table*}[tb]
\caption{Metrics comparing two sets of qrels via two labelling strategies. $|S_{gt}|$ / $|NS_{gt}|$ is presented under the dataset name.}
\centering
\begin{tabular}{l|c|ccccccccccc}
\toprule
Dataset            & Qrels         & $\kappa$ & $\tau$ & $\Delta$ Sens. & \precone{}  & \recone{} & \prectwo{} & \rectwo{} & BAC    & MCC & FP & FN    \\ \midrule
\multirow{2}{*}{\shortstack{DL19\\173/465 sig.}} & Popularity-Biased  &  0.176   & 0.230 & 0.366 & 0.000 & 0.000 & 0.626      & 0.990        & 0.495 & -0.062 & 2 & 176  \\
& Zero-Shot               & 0.125   & 0.845 & 0.191  & 1.000 & 0.500 & 0.763      & 1.000        & 0.750 & 0.618   & 0 & 85\\ \midrule
\multirow{2}{*}{\shortstack{DL20 \\$714/1711$ sig.}} & Popularity-Biased & 0.220   & 0.528 & 0.227         & 0.994 & 0.455 & 0.718 & 0.998   & 0.726 & 0.568 & 2 & 77  \\
& Zero-Shot &  0.141   & 0.855 & 0.219         & 0.985 & 0.459 & 0.725      & 0.995       & 0.727 & 0.568  & 3 & 384\\
  \bottomrule
\end{tabular}%
\label{tab:table_1}
\vspace{-4mm}
\end{table*}

\section{Experimental Setup} \label{sec:exp_setup}
We investigate hypothesis testing errors through two research questions: \textbf{RQ1}: Does measuring the propensity of Type II errors provide insights into the discriminative power of qrels? \textbf{RQ2}: Can balanced classification metrics be used to provide an overall view of the discriminative power of qrels? 

\pageenlarge{2}
\looseness -1
\noindent\textbf{Datasets} We use the TREC Deep Learning 2019 (DL19)~\cite{dl19} and 2020 (DL20)~\cite{dl20} datasets\crc{, which are commonly used in the recent literature.} To rank systems and measure significant differences we use nDCG@10 scores of the official runs submitted to TREC (32 for DL19 and 59 for DL20).

\noindent\textbf{Metrics}
\looseness -1
Following~\cite{auto_eval_soboroff,thomas_predict,umbrella}, we use Cohen's $\kappa$ to measure the agreement of labels between ground truth and candidate qrels and Kendall's $\tau$ to measure system ranking correlation. The sensitivity (sens.)~\cite{perspectives} metric is defined as $\frac{\text{\# distinguished pairs}}{\text{\# total pairs}}$, where a distinguished pair of systems is one with a significant difference. We report $\Delta$ sensitivity between ground-truth and candidate qrels. \jm{We use \precone{}, \prectwo{}, \recone{}, and \rectwo{} to measure discriminative power.} High \precone{} and \prectwo{} values indicate few Type I and II errors respectively and high \prectwo{} and \rectwo{} indicate high coverage. The balanced classification metrics investigated are Balanced Accuracy (BAC)~\cite{bac} and Matthews Correlation Coefficient (MCC)~\cite{mcc}. We correct for multiple comparisons~\cite{multiple_comparisons_two} using the Paired Randomised Tukey HSD test~\cite{tukey_hsd}.

\section{Results and Discussion}
\looseness -1
\textbf{RQ1} investigates quantifying Type II errors by measuring the precision and recall of non-significant differences. We compare qrels using $\Delta$ sensitivity, \precone{}, \recone{}, \prectwo{}, and \rectwo{} to answer \textbf{RQ1}. Each row of Table~\ref{tab:table_1} represents candidate qrels for a dataset, while columns show the metrics that are used to compare the qrels. We can see from Table~\ref{tab:table_1} that, for DL19, the zero-shot labeller is less likely to make Type II errors than the popularity-biased labeller. The zero-shot labeller qrels have 22.0\% higher \prectwo{} than the popularity-biased labeller qrels, indicating that more of the truly non-significant differences are identified and therefore fewer Type II errors are made. We gain extra information about hypothesis testing errors that are not possible otherwise by looking at the \prectwo{} measure that we propose to use.

\begin{table}[b]
\vspace{-4mm}
\caption{Comparison of the impact of the number of queries on discriminative power metrics. Randomly selected subset of queries from the DL20 Popularity-Biased Labeller qrels.}
\label{tab:num_queries}
\vspace{-2mm}
\resizebox{0.47\textwidth}{!}{%
\begin{tabular}{@{}l|cccccccc@{}}
\toprule
\#Queries & \precone{} & \recone{} & \multicolumn{1}{c}{\prectwo{}} & \multicolumn{1}{c}{\rectwo{}} & \multicolumn{1}{c}{TP} & \multicolumn{1}{c}{TN} & \multicolumn{1}{c}{FP} & \multicolumn{1}{c}{FN} \\ \midrule
10 & 0.9248                     & 0.3229                     & 0.2459                              & 0.8938                           & 443                    & 303                    & 36                     & 929                    \\
30 & \multicolumn{1}{r}{0.8844} & \multicolumn{1}{r}{0.4461} & 0.2542                              & 0.7640                           & 612                    & 259                    & 80                     & 760                    \\ \bottomrule
\end{tabular}%
}
\end{table}

\begin{figure}[tb]
    \centering
    \includegraphics[width=\linewidth]{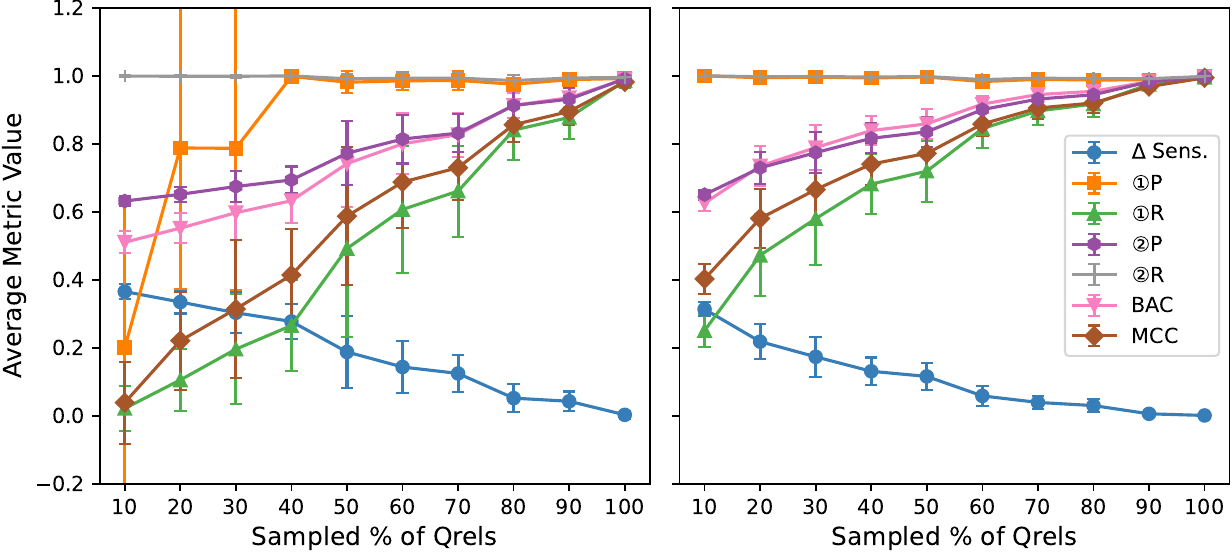}
    \caption{Effect of sampling qrels on classification metrics. Left: DL19. Right: DL20.}
    \label{fig:side_by_side}
    \vspace{-5mm}
\end{figure}

Examining the popularity-biased labeller qrels using $\Delta$ sens., we see that there is a large difference in the number of distinguished pairs of systems identified when comparing with the ground truth qrels. However, $\Delta$ sens. does not indicate if Type I or Type II errors occur, as it simply counts the distinguished pairs that are identified, regardless of any errors made. Turning to \precone{} and \recone{}, no conclusions can be drawn as all of the significant differences identified are incorrect and none of the true significant differences are identified, hence \precone{} = \recone{} = 0. The \precone{} and \recone{} metrics proposed by Otero et al.~\cite{how_discrim} do not enable us to draw any conclusions on the discriminative power of the popularity-biased labelled qrels. Indeed, we cannot conclude on Type I errors in this case, as it is possible that no distinguished pairs are identified at all. On the other hand, if we look at \prectwo{} and \crc{\rectwo{}}, we can gain information on the discriminative power of the popularity-biased labeller that was not available before.\pageenlarge{2} We can see that $\sim$63\% of pairs of systems with non-significant differences identified are correct (0.626 \prectwo{}) and that almost all the non-significant differences are identified. Therefore, we see that $\sim$36\% of non-significant differences identified are in fact Type II errors. Using the \prectwo{} and \rectwo{} metrics that we propose has allowed us to identify Type II errors that have occurred.
Moreover, using \prectwo{} and \rectwo{} provides a more interpretable quantification of the propensity of Type II errors that occur when using a set of candidate qrels than \precone{} and \recone{}. Making a hypothesis testing error changes the numerator in both families of measure and therefore Type II errors are implicitly considered by \precone{} and \recone{}. However, \prectwo{} and \rectwo{} specifically model Type II errors and consequently provide a more interpretable measure of hypothesis testing errors.

\vspace{-2mm}
\observation{\prectwo{} gives insights into the rate of Type II errors that occur that cannot be observed when using \precone{} and \recone{}.}
\vspace{-2mm}

To illustrate this interpretability, we measure the discriminative power of two sets of qrels. We select two random subsets of queries from the Popularity-Biased Labeller qrels of DL20 as our candidate qrels. Table~\ref{tab:num_queries} shows discriminative power metrics and the four confusion matrix outcomes for subsets of 10 and 30 queries. One would expect that evaluation over 10 queries would result in more Type II errors as the statistical test has less power~\cite{stat_pow_1,stat_pow_2}. Indeed, we see more false negatives with 10 queries than we do with 30 queries (920 vs. 760), indicating more Type II errors. By observing \precone{} and \recone{}, we see that more false positives (Type I errors) occur when using a smaller subset of queries. We cannot, however, gather interpretable information about which set of qrels results in more Type II errors. When a Type II error occurs, the numerator of \precone{} and \recone{} necessarily decreases. However, the propensity of Type II errors is not easy to interpret as the outcome of the statistical test may be correct. On the other hand, if we look at \prectwo{}, for example, we can see that fewer Type II errors occur when using a subset of 30 queries than occur when using 10 queries as there is an increase in \precone{} (0.2459 $\rightarrow$ 0.2542). We can therefore conclude by answering \textbf{RQ1}; measuring the precision and recall of non-significant differences being identified allows us to gain further insights into the Type II statistical errors that can occur when comparing qrels, providing further insights into their discriminative power.

\pageenlarge{2}
\vspace{-2mm}
\observation{\prectwo{} is more interpretable than \precone{} regarding Type II errors.}
\vspace{-2mm}

\looseness -1
\textbf{RQ2} examines if a single, balanced, classification metric can distil the discriminative power of qrels into one number. Having a single metric quantifying Type I and II errors is useful when creating new qrels. We investigate BAC and MCC as our measures. Figure~\ref{fig:side_by_side} shows the percentage of relevant documents sampled on the x-axis and the average performance of 10 \jm{sampling repetitions} on the y-axis, with error bars showing the variance. We can see that BAC is robust to large changes in \jm{the rates of Type I and Type II errors}. Moving from sampling 30\% of relevant qrels from DL19 to 40\%  results in a large increase in \precone{}, whereas other metrics have a more steady increase. BAC reflects this with a steady and proportionate increase.

\vspace{-2mm}
\observation{Balanced classification metrics provide an overall view of the hypothesis testing errors when comparing IR systems.}
\vspace{-2mm}

On DL20, we see that BAC closely aligns with \prectwo{}, as \precone{} and \rectwo{} $\approx$ 1.0 for all sampling rates and \recone{} starts high and increases steadily. MCC demonstrates another robust reflection of the discriminative power of a set of qrels as both MCC and BAC take into account true and false positives and negatives and consequently, both Type 1 and Type 2 errors are reflected. \jm{Whilst both the tested balanced metrics condense discriminative power into a single number, since MCC is in the range \gm{[-1,+1]} \gm{with a random performance receiving a MCC score of 0,} it provides a \gm{more} interpretable measure of how well the outcomes of the significance tests both agree and disagree. As such, it can be recommended as a single interpretable measure of the discriminative power of qrels, in answer to \textbf{RQ2.}}

\section{Conclusions}
In this work, we argued that Type II hypothesis testing errors are important to measure when comparing different qrels, something previously neglected. We, therefore, investigated the quantification of Type II errors using precision and recall of non-significant differences using LLM-generated and subsampled qrels. Additionally, we observe that balanced classification metrics give an overall view of the discriminative power of qrels as they consider Type I and Type II errors in one measure. The information gained by measuring statistical errors using appropriate metrics can lead to more robust IR evaluation and accurate conclusions.


\balance

\end{document}